# Direct evidence for superconductivity in the organic charge density-wave compound α-(BEDT-TTF)$_2$KHg(SCN)$_4$ under hydrostatic pressure

D. Andres[1], M. V. Kartsovnik[1], W. Biberacher[1], K. Neumaier[1], H. Müller[2]

[1]Walther-Meissner-Institut, Bayerische Akademie der Wissenschaften, D-85748 Garching, Germany
[2]European Synchrotron Radiation Facility, F-38043 Grenoble, France

The layered organic metal α-(BEDT-TTF)$_2$KHg(SCN)$_4$ possesses a strongly anisotropic electron system with coexisting quasi-one-dimensional (Q1D) and quasi-two-dimensional (Q2D) conducting bands. Numerous experiments point to a nesting instability of the Q1D parts of the electronic bands causing a formation of a charge density-wave (CDW) at ≈ 8 K ( for a review see, e.g., [1,2] and references therein). The Q2D band remains metallic that determines a decreasing resistance with cooling down to lowest temperatures. At ambient pressure, the resistance decrease has been found to significantly accelerate below ≅ 300 mK [3]. This acceleration is most likely of a superconducting (SC) origin as it follows from its dependence on the current level and weak magnetic field [3]. The SC transition is, however, incomplete that was interpreted [3] in terms of the proximity of the in-plane sheet resistance to the critical value $h/4e^2$ of a superconductor-insulator transition in disordered two-dimensional superconductors [4]. The isomorphous salt α-(BEDT-TTF)$_2$NH$_4$Hg(SCN)$_4$ does not undergo the CDW transition but instead becomes superconducting at $T_c$ ≈ 1 K [5]. The rather sharp bulk SC transition in that compound was thus explained [3] by the sheet resistance far below $h/4e^2$.

A hydrostatic pressure suppresses the CDW state in α-(BEDT-TTF)$_2$KHg(SCN)$_4$, re-establishing the normal metallic (NM) state above $P_0$ ≈ 2.5 kbar [2]. In the above discussion of incomplete superconductivity it is assumed that the density wave increases the sheet resistance. Thus, one might expect that the suppression of the CDW by pressure would lead to an enhancement of superconductivity. However, no superconductivity under hydrostatic pressure has been observed so far. Here, we present direct evidence of a SC state existing below 300 mK under quasi-hydrostatic pressure. The character of the SC transition drastically changes, with changing pressure, upon crossing the CDW/NM phase boundary.

The data presented here were taken from interlayer resistance measurements on two different samples; pressure was applied using the conventional clamp cell technique. The cell was mounted on a dilution refrigerator allowing the sample to be cooled down to ≈ 20 mK. To avoid overheating the sample current was kept ≤ 100 nA. Thus overheating was estimated as < 5 mK at $T$ = 20 mK.

Fig.1 shows temperature dependences of the resistance of sample #1 at various pressures. As can be seen in the NM region (Fig.1a), i.e. at $P > P_0$, relatively sharp SC transitions are observed. The transition temperatures $T_c$, extracted from the inflection point of the transition, are presented as filled circles in Fig. 2. $T_c$ clearly decreases with increasing pressure, a scenario commonly observed in the NM states of organic metals [8]. On lowering the pressure below $P_0$ (Fig.1b), the SC transition becomes strongly influenced by the presence of the density wave that can be described as follows: (i) The transition itself changes, becoming broader the better the nesting of the density wave is. Due to the broadening, an exact determination of $T_c$ cannot be done at $P<P_0$. We therefore determine on- and offset transition temperatures as shown in Figs. 1b and 2. (ii) Within the transition several distinct steps emerge. (iii) Already at temperatures far above the step features the resistance starts to decrease in a SC-like manner. The starting temperature (not shown) is found to be slightly pressure dependent: 250 mK at 2.5 kbar and

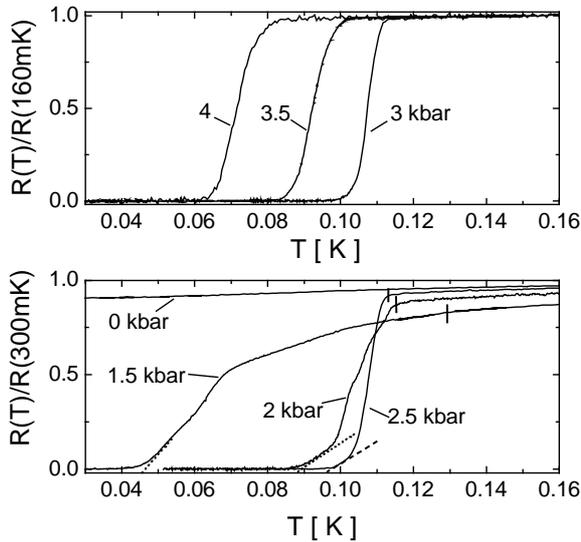
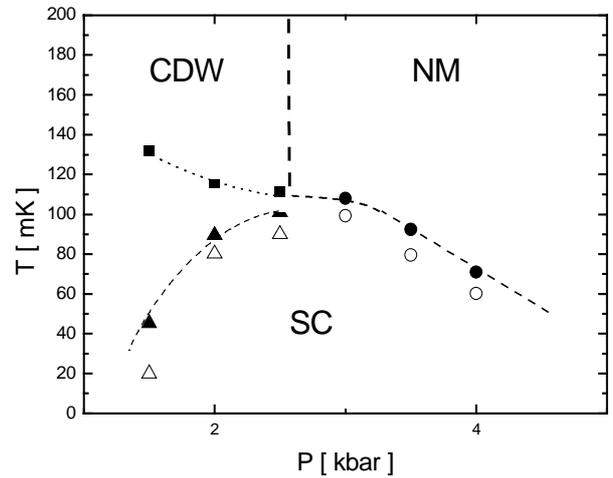

**Figure 1.** SC transitions in the temperature dependence of the interplane resistance at (a) P > 2.5 kbar and (b) P ≤ 2.5 kbar. In (b) an onset of the transition is determined by a step-like change of the slope (vertical dashes) and an offset by linear extrapolations to zero-resistance (dotted lines).

**Figure 2.** Low-temperature part of the $T$-$P$ phase diagram: circles show inflection points of the SC transition in the M region; squares/triangles show the onsets/offsets of the transiton in the CDW region. Filled and open symbols correspond to sample #1 and #2, respectively. Dashed lines are guides for the eye.

300 mK at 2 kbar and 0 kbar. This decrease is easily suppressed by a small magnetic field and depends on current as already reported for $P = 0$ kbar [3], suggesting the presence of small superconducting regions or filaments.

Obviously the sharp transitions at $P \geq 2.5$ kbar are far below the proposed ambient-pressure $T_c$, i.e. 300 mK. The incomplete superconductivity within the CDW state therefore can hardly be attributed to the proximity of an insulator transition, since in the latter model $T_c$ is expected to increase with decreasing the sheet resistance. Further, $dT_c/dP$ in the M state of the title compound is found to be ≈ 30 mK/kbar, that is nearly an order of magnitude smaller than observed in $\alpha$-(BEDT-TTF)$_2$NH$_4$Hg(SCN)$_4$ [5]. This might be due to different parts of the Fermi surface contributing to superconductivity in both compounds. Therefore a direct comparison of the SC properties is likely inappropriate.

The observed sample dependence of the transition temperatures (Fig. 2) points to a possibly non-pure s-wave nature of the superconducting order parameter, as it already has been suggested in some other BEDT-TTF based superconductors [5]. We expect crystal defects or impurities to have a big effect on $T_c$, since in both samples the crystal quality, extracted from the residual resistance ratios or from the Shubnikov-de Haas oscillations of the Q2D band, appeared to be similarly high. Remarkably, within the CDW state of our compound the sample dependence of the transition points is found to become even stronger (see 1.5 kbar in Fig.2). Thus the additional influence of the CDW on superconductivity is also most likely dependent on impurities or defects. Indeed, such a dependence is proposed for a CDW and superconductivity coexisting on an imperfectly nested Fermi surface [6]. $T_c$ in this model is proposed to decrease the better the nesting conditions are. If this is the case here, there would still remain the question why such a strong broadening of the transition in the CDW state occurs. To clarify the situation it would be in particularly helpful to perform measurements of the inplane resistance.

**References**

[1] Kartsovnik M. V., Andres D., Biberacher W., Christ P., Togonidze T., Steep E., Balthes E., Müller H., Kushch N. D., *Synth. Met.* **120** (2001) 687.
[2] Andres D., Kartsovnik M., Biberacher W., Weiss H., Balthes E., Müller H., Kushch N., *Phys. Rev. B* **64** (2001) 161104(R).
[3] Ito H., Kartsovnik M. V., Ishimoto H., Kono K., Mori H., Kushch N. D., Saito G., Ishiguro T., Tanaka S., *Synth. Met.* **70** (1995) 899.
[4] Fisher M., *Phys. Rev. Lett.* **65** (1990) 923.
[5] Ishiguro T., Yamaji K., Saito G., Organic Superconductors (Springer-Verlag, Berlin, Heidelberg 1998).
[6] Gabovich A. M., Voitenko A. I., *J. Low. Temp. Phys.* **26** (2000) 305.